\documentclass[aps,pre,eqsecnum]{revtex4}
\usepackage{amsfonts,amssymb,amsmath,bm}
\usepackage[dvipdf]{color}
\usepackage{graphicx}

\definecolor{RED}{rgb}{0.8,0.,0.}
\definecolor{BLUE}{rgb}{0.,0.,0.6}

\begin{document}

\title{Tail-Constraining Stochastic Linear-Quadratic Control:\\
Large Deviation and Statistical Physics Approach}

\author{Michael Chertkov $^{(1)}$, Igor Kolokolov $^{(2,1)}$, and Vladimir Lebedev $^{(2,1)}$\\
Center for Nonlinear Studies and Theoretical Division at LANL,\\
\& New Mexico Consortium, Los Alamos, NM, USA $^{(1)}$\\
Landau Institute for Theoretical Physics, Moscow, Russia $^{(2)}$}

\begin{abstract}
Standard definition of the stochastic Risk-Sensitive Linear-Quadratic (RS-LQ) control depends on the risk parameter, which is normally left to be set exogenously. We reconsider the classical approach and suggest two alternatives resolving the spurious freedom naturally. One approach consists in seeking for the minimum of the tail of the Probability Distribution Function (PDF) of the cost functional at some large fixed value. Another option suggests to minimize the expectation value of the cost functional under constraint on the value of the PDF tail. Under assumption of the resulting control stability, both problems are reduced to static optimizations over stationary control matrix. The solutions are illustrated on the examples of scalar and 1d chain (string) systems. Large Deviation self-similar asymptotic of the cost functional PDF is analyzed.
\end{abstract}

\maketitle

\section{Introduction}
\label{sec:Intro}

Stochastic differential equations are used both in control \cite{49Wei,Kalman1960,60Kal,60Str,61KB,63Kal,74Kai} and statistical physics \cite{05Ein,06Smo,57Ula,92Kam,04Gar} to state the problems. The two fields also use similar mathematical methods to analyze these equations. However, and in spite of the commonalities, there were relatively few overlaps between the disciplines in the past, even though the communications between two communities improved in the recent years.  Some new areas in control, for example stochastic path integral control \cite{05Kap,11Kap,10BWK,Dj11}, have emerged influenced by analogies, intuition and advances in statisical/theoretical physics.  Vice versa, many practical experimental problems in physics, chemistry and biology dealing with relatively small systems (polymers, membranes, etc), which are driven and experience significant thermal fluctuations, can now be analyzed and manipulated/controled with accuracy and quality unheard of in the past, see for example \cite{11Jar,05BLR}. Besides, approaches from both control theory and statistical physics started to be applied to large natural and engineered networks,  like chemical, bio-chemical and queuing networks \cite{97MA,07SS,10CCGT,11CCS}. Dynamics over these networks is described by stochastic differential equations, the networks have enough of control knobs, and they function under significant fluctuations which need to be controlled to prevent rare but potentially devastating failures. Related setting of stochastic optimization, i.e. optimization posed under uncertainty, has also came recently in the spot light of statistical physics inspired algorithms and approaches \cite{ABRZ2011}.
Convergence of these are related ideas motivated the manuscript,  where we discuss analysis and control of rare events in the simplest possible,  but practically rather widespread universal and general, linear setting. We realize that the general topic of linear control is well studied and many (if not all) possible questions, e.g. related to proper way of accounting for risk (rare events),  were discussed in the field in the past. In spite of that,  we still hope that this manuscript may also be useful not only to physicists, who may wish to explore new and largely unusual (in physics) formulations, but also to control theorists.

Consider first order (in time derivatives) stochastic linear dynamics of a vector $x=(x_i|i=1,\cdots,N)$ over time interval $t'\in[t;T]$
\begin{eqnarray}
\frac{d}{dt'}x=Ax+Bu+\xi(t'),
\label{stoch}
\end{eqnarray}
where $A$ and $B$ are constant matrices; $u(t')$ is the control vector applied at the moment of time $t'$; and $\{\xi\}=(\xi(t')|t'\in[t;T])$ is the zero mean,  short-correlated noise with covariance $V$
\begin{eqnarray}
\langle \xi_i(t')\rangle=0,\quad \langle \xi_i(t')\xi_j(t'')\rangle =\delta(t'-t'')V_{ij},\quad i,j=1,\cdots,N
\label{xi}
\end{eqnarray}
where one utilizes "statistical physics" notations for the expectation value (average) over noise, $\langle\cdots\rangle$. Here in Eq.~(\ref{xi}) and below the averages are over multiple possible realizations of the noise, each generating a new trajectory of the system, $\{x\}=(x(t')|t'\in[t;T])$, under given control $\{u\}=(u(t')|t'\in[t;T])$. The Eq.~(\ref{stoch}) is causal, thus assuming retarded (Stratonovich) regularization of the noise on the right-hand-side of the discreet version of Eq.~(\ref{stoch}). The physical meaning of the vectors and matrices in Eq.~(\ref{stoch}) is as follows. $A$ is the matrix explaining stretching, shearing and rotation of the system trajectory in the $N$-dimensional space if the control and external noise would not be applied. Matrix $B$ describes possible limitations on the degrees of freedom in the system one can control. To simplify notations we consider signal, control and noise vectors having the same dimension, $N$, where thus $B$ is quadratic. The setting of Eqs.~(\ref{stoch},\ref{xi}) is classic one in the control theory. It describes the so-called Linear-Quadratic (LQ) stochastic control problem, which was introduced in \cite{Kalman1960,60Kal,61KB,63Kal} and became foundational for the control theory as a field, see e.g. \cite{Willems1971,Athans1971} and references therein.  In the classical formulation one seeks to solve the following optimization, $t\in[0;T]$:
\begin{eqnarray}
&& \min_{\{u\}} \Biggl\langle J(t;T;\{u\}, \{x\})\Biggr\rangle,
\label{LQ}\\
&& \label{cost} J(t;T;\{u\}, \{x\})\equiv\frac{1}{2}x^*(T)F x(T)
+\frac{1}{2}\int_t^T dt' \left( x^*(t')Q x(t')+u^*(t')R u(t')\right),\nonumber
\end{eqnarray}
where $Q,R$ and $F$ are pre-defined stationary (time independent) symmetric positive matrices and one uses the super-script asterisk, $*$, to mark transposition. $J(t;T;\{u\}, \{x\})$ (later on, and when it is not confusing, we will use the shortcut notation $J$) is a scalar quadratic cost functional of the state vector  $\{x\}=(x(t')|t'\in[t;T])$, and the control vector, $\{u\}=(u(t')|t'\in[t;T])$ evaluated for all intermediate times $t'$ from the $[0;T]$ interval. Here in Eqs.~(\ref{LQ}) (and everywhere below in the manuscript) the average over noise $\{\xi\}$ includes conditioning to Eq.~(\ref{LQ}),  i.e. $\{x\}$ is dependent on realization of the noise, $\{\xi\}$, and on the control, $\{u\}$,  according to Eq.~(\ref{stoch}). It is assumed that the stochastic LQ control is evaluated off-line, i.e. the optimal solution $u_*(t;x(t))$ of Eq.~(\ref{LQ}) is computed and saved prior to executing actual experiment for any initial condition $x(t)$ at any $t$.  Then in the course of the actual experiment (execution of the dynamics) $x(t)$ is measured at any time $t$ and respective $u_*(t;x(t))$ is applied. (When observation of $x(t)$ is partial and noisy one needs to generalize the stochastic LQ control,  for example considering the stochastic Linear Quadratic Gaussian (LQG) control, see e.g. \cite{Athans1971} for details.)  We also assume (and the details will be clarified below) that the optimal control succeeds, i.e. the systems stabilizes and $J$ does not grow with $T$ faster than linearly.

An unfortunate caveat of the LQ setting (\ref{LQ}) is in the lack of fluctuations control: even though the LQ solution is optimal in terms of minimizing the expectation value of the cost functional it may generate very significant fluctuations when it comes to analysis of the ${\cal J}\gg \langle J\rangle$ tail of the Probability Distribution Function, ${\cal P}({\cal J})$, of the cumulative cost ${\cal J}(T;x(0))\equiv J(0;T;\{u_*\}, \{x\})$. Stochastic Risk Sensitive LQ (RS-LQ) scheme \cite{Jacobson1973,Whittle1981,Whittle1986} was introduced to improve control of the abnormal fluctuations of $J$. RS-LQ constitutes the following generalization of the LQ scheme (\ref{LQ})
\begin{eqnarray}
&& \max_{\{u\}} \langle \exp\left(-\theta J\right)\rangle,
\label{RS-LQ}
\end{eqnarray}
where $\theta$ is a pre-defined parameter. Intuitively one relates the case of positive $\theta$ to a risk-avert optimum.
It is assumed within the standard RS-LQ scheme that $\theta$ is fine tuned by some additional considerations. Note that, as shown in \cite{Cb1988}, the RS-LQ control is also ultimately related to the so-called $H_\infty$-norm robust control.
 (See also \cite{00PJD} for further discussion of the relation.)

In this paper we analyze two natural modifications of the stochastic RS-LQ control. The two schemes can both be interpreted in terms of the RS-LQ approach supplemented by an additional optimization over $\theta$. Our first,  \emph{Tail-Optimum} (TO), scheme consists in the following modification of the LQ (\ref{LQ}) and RS-LQ (\ref{RS-LQ}) ones
\begin{eqnarray}
\min_{\{u\}} {\cal P}(J=j\cdot(T-t)|\{u\}).
\label{TO-LQ}
\end{eqnarray}
In words, the TO-LQ control minimizes (at any time $t$ and given the current observation $x(t)$) the probability of the current value of the cost functional $J(t;T;\{u\}, \{x\})$ evaluated at a predefined value, $j\cdot(T-t)$, where thus $j$ is the only external parameter left in the formulation. Another strategy, which we call {\it Chance-Constrained} LQ (CC-LQ), in reference to similar formulations in optimization theory \cite{58CCS,65MW,09BEN}, consists in minimizing the mean of the cost functional under condition that the tail probability evaluated at  $j\cdot(T-t)$ does not exceed the prescribed threshold value $\varepsilon(t;T)$
\begin{eqnarray}
&\min\limits_{\{u\}} &\langle J\rangle
\label{CC-LQ}\\
&& \mbox{s.t. } {\cal P}(J=j\cdot(T-t)|\{u\})\leq \varepsilon(t;T).
\nonumber
\end{eqnarray}

Main objectives, and consequently results of this study, are\\
$\bullet$ {\it To extend the asymptotic, $T\to\infty$, approach, developed in the past for LQ (\ref{LQ}) and RS-LQ (\ref{RS-LQ}) optimal controls to the new TO-LQ (\ref{TO-LQ}) and CC-LQ (\ref{CC-LQ}) optimal settings.} At $T\to\infty$ the optimal control takes the following universal linear in $x$ form
\begin{eqnarray}
u_*(t;x)=-Kx, \label{u-opt}
\end{eqnarray}
where $K$ is $t$-independent but model dependent matrix.
The condition of the system stability,  intuitively translating into the expectation that ${\cal J}$ grows not faster than linearly with $T$, naturally requires that all the eigenvalues of the stability matrix, $\mu=BK-A$,  have positive real part.
The linearity of the optimal control (\ref{u-opt}) in $x$ is a direct consequence of the linearity of the initial dynamical system.
Time-indepedence and  initial condition-independence of the optimal control (\ref{u-opt}) are asymptotic: they are
achieved at $T\gg\tau_*$, where $\tau_*$ can be estimated  as the inverse of the absolute value of the $(BK-A)$'s
eignevalue with the smallest real part.
System of algebraic equations defining $K$ implicitly for the TO-LQ and CC-LQ cases are presented and then juxtaposed against the previously analyzed cases of the LS and RS-LQ controls.
(See Eqs.~(\ref{RS-opt},\ref{TO-opt},\ref{CC-opt}).) Finding optimal control is reduced to optimization over time-independent, $K$.
The resulting dependencies are homogeneous in time, with $t$ and $T$ always enter in the $T-t$ combination.
(This also simplifies the analysis allowing to set $t=0$.) \\
$\bullet$ {\it To analyze statistics of the optimal cost functional, ${\cal J}$, in the Large Deviation (LD) regime,  i.e. at large but finite $T$.} We show that in the stable regime the PDF of ${\cal J}$ attains the following universal LD form
\begin{eqnarray}
\log {\cal P}({\cal J})\sim -T {\cal S}({\cal J}/T),
\label{LD}
\end{eqnarray}
where the LD function, ${\cal S}(j)$, is a convex function of its argument found implicitly (in a closed algebraic form, which may or may not yield an efficient algorithm) for the four cases (of LS, RS-LQ, TO-LQ and CC-LQ controls) considered. The LD function shows a universal, ${\cal S}(j)\to a j$, tail at large (i.e. larger than typical) $j$,  where the value of positive $a$ depends on the model.
This suggests,  in particular, that it is natural to choose in the CC optimization (\ref{CC-LQ}), $\varepsilon(t;T)=\exp(-c(T-t))$, for the threshold, with $c$ been a constant. To derive compact algebraic expressions for the LD function we, first, analyze the generating function of ${\cal J}$ evaluated at linear $u$ parameterized by $K$ as in Eq.~(\ref{u-opt}),
\begin{eqnarray}
&& {\cal Z}(\theta;K)\equiv \langle\exp\left(-\theta {\cal J}\right)\rangle_*,
\label{Z}
\end{eqnarray}
then express the optimization/control objective as a convolution of the integral or differential operator/kernel in $\theta$ (the choice will depend on the model) and ${\cal Z}(\theta;K)$, and finally evaluate optimization over $K$ in the asymptotic LD approximation.
Here in Eq.~(\ref{Z}) the low asterisk mark $*$  in the expectation/average (over noise and constrained to Eq.~(\ref{stoch})) indicates that the control vector is taken in the form of the linear ansatz, $u\to - K x$, where $K$ is left yet undetermined.

The remainder of the manuscript is organized as follows. We start discussing the deterministic case (of zero noise) in Section \ref{sec:Det}. This regime is of interest for two reasons.  First, in the asymptotic of zero noise the four, generally different, control schemes become equivalent. Besides,  and as well known from the classical papers \cite{Kalman1960,Willems1971,Athans1971}, optimal control in the bare LQ case (correspondent to minimization of the cost function average) is not sensitive (and thus independent of) the level of the noise. Section \ref{sec:Z} is devoted to analysis of the generating function (\ref{Z}), the average value of the cost function and the tail of the cost function distribution restricted to yet unspecified value of $K$. Optimization over $K$, resulting in the known RS-LQ optimal relations and also derivation of the new optimal relations for $K$ in the TO-LQ and CC-LQ cases, is discussed in Section \ref{sec:Opt}. We describe and compare asymptotic Large Deviation forms of the cost function PDF, ${\cal P}({\cal J})$, in the optimal regimes. In this and preceding Section we also discuss many times the illustrative "scalar" example,  where $x$ and $u$ are scalars. An infinite system example, of a "string" formed from a linear 1d chain, is discussed in Section \ref{sec:String}. We conclude and discuss related future challenges in Section \ref{sec:Con}.

\section{Deterministic Case and LQ-optimal control}
\label{sec:Det}

We start this Section from a disclaimer: all results reported here are classical, described in \cite{Kalman1960,Willems1971,Athans1971,Jacobson1973,Whittle1981,Whittle1986,Cb1988} and latter papers and books, see e.g. \cite{02FPE,96ZDG,09Hes}. We present it here only for making the whole story of the manuscript self-explanatory and coherent.

When the noise is ignored, Eq.~(\ref{stoch}) should be considered as a deterministic constraint, reducing any of the optimal control schemes (\ref{LQ},\ref{RS-LQ},\ref{TO-LQ},\ref{CC-LQ}) to a simple variation of the cost functional (\ref{cost}) over $u$. Using the standard variational technique with a time dependent Lagrangian multiplier for the constraint, and then excluding the multiplier one derives the equation
\begin{eqnarray}
&& \frac{d}{dt'}u^*+u^*RB^{-1}ABR^{-1}=x^*QBR^{-1},
\label{u_det}
\end{eqnarray}
which should be supplied by the boundary condition (also following from the variation), $u^*(T)+x^*(T)FBR^{-1}=0$.
(Let us remind that we choose the notations where the dimensionality of $u$ coincides with the dimensionality of $x$. We also assume that inverses of all the matrices involved in the formulation are well defined.  This assumption is not critical and is made here only to simplify the notations.  In the general case when some of the matrices,  in particular $R$, are not full rank,  one can generalize the formulas properly,  using a proper notion of the pseudo-inverse.) Substituting, $u=-R^{-1}B^*\Pi x$, in Eq.~(\ref{u_det}) one arrives at the following equation for $\Pi$
\begin{eqnarray}
\frac{d}{dt'}\Pi+\Pi A+A^*\Pi-\Pi B R^{-1}B^*\Pi+Q=0.
\label{Pi}
\end{eqnarray}
with the boundary condition $\Pi(T)=F$. Eq.~(\ref{Pi}), solved backwards in time, results in
$\Pi(t)$ and then, $u_*(t;x)=-R^{-1}B^*\Pi(t)x=-Kx$.

To gain a qualitative understanding of the backwards in time dynamics of $\Pi$,  let us briefly discuss the simplest possible case with all the matrixes entering Eq.~(\ref{Pi}) replaced by scalars, then yielding the following analytic solution for the optimal $K$
\begin{eqnarray}
K=\frac{A-\sqrt{A^2+\frac{QB^2}{R}}\left(\tanh\left(\sqrt{A^2+\frac{QB^2}{R}}(t-T_0)\right)\right)^{\pm 1}}{B},
\label{K_1_no_noise}
\end{eqnarray}
where $T_0$ and $\pm 1$ are chosen to satisfy the boundary condition, $K(T)=BF/R$. When $T\gg \tau=1/\sqrt{A^2+QB^2/R}$,  the backwards in time dynamics saturates (after a short $\sim \tau$ transient) to a $F$-independent constant, resulting from replacing $\tanh$ in Eq.~(\ref{K_1_no_noise}) by $-1$.
Therefore,  in the stationary regime, $T\to\infty$ the optimal control is with the constant in time,  frozen $K$. One also finds that the optimal control in the one dimensional deterministic case is always stable, $\mu=KB-A>0$.

Returning to the general (finite vector) case one concludes that when $T$ is sufficiently large the optimal control is of the form described by Eq.~(\ref{u-opt}), i.e. it is linear in $x$ and asymptotically  time independent, with $K=R^{-1}B^*\Pi_0$ where $\Pi_0$ solves Eq.~(\ref{Pi}) with the first term replaced by zero. It is well known in the control theory that (under some standard common sense assumptions on $B$ and $R$ matrices) stable solution of the system of the algebraic Riccati equations is unique and moreover it can be  found efficiently. (See e.g. Chapter 12 of Sec\cite{98Zho} and references therein.)

Let us now discuss the bare LQ control,  now in the presence of the noise. Since Eq.~(\ref{stoch}) is linear, one can naturally split the full solution into a sum, $x=x_1+x_2$,  where $x_1$ satisfies Eq.~(\ref{stoch}) without noise and it is equivalent to the noise-less solution,  just discussed in this Section.  Then, the second term satisfies, $dx_2/dt'=A x_2+\xi$, with $x_2(t)=0$. However,  since the noise is zero mean, $\langle\xi\rangle=0$,
$x_2$ is zero mean too, i.e. $\langle x_2\rangle=0$. Next,  let us analyze the split of term in $\langle J\rangle$,  which is the optimization objective of the LQ scheme. Since, $x_1$ and $x_2$ are independent (by construction) and because $x_2$ is zero mean, $\langle J\rangle$, splits into two terms, $\langle J_1\rangle+\langle J_2\rangle$, each dependent on $x_1$ and $x_2$ vectors only. $\langle J_1\rangle$ is simply equivalent to $J$ analyzed above in the deterministic case, while $\langle J_2\rangle$ is $u$-independent,  thus not contributing the optimization at all.
To summarize, the LQ optimal control is not sensitive to the noise and it is thus equivalent to the deterministic (noiseless) case described above.

\section{Generating Function}
\label{sec:Z}

Consider the Generating Function (GF), ${\cal Z}(\theta;K)$, defined by Eq.~(\ref{Z}).  ${\cal Z}(\theta;K)$ is of an obvious relevance to the RS-LQ scheme,  but it is also useful for analysis of other schemes as well,  because of the following (Laplace transform) relation to the PDF of ${\cal J}$:
\begin{equation}
 {\cal Z}(\theta;K)=\int\limits_0^\infty d{\cal J}\
 \exp(-\theta {\cal J}) {\cal P}_*({\cal J}),
 \label{Laplace1}
 \end{equation}
where (as before) the  asterisk in the sub-script indicates that the PDF was evaluated at $u=Kx$,  with $K$ being yet undefined constant matrix. The inverse of Eq.~(\ref{Laplace1}) is
 \begin{equation}
 {\cal P}_*({\cal J})=\int\limits_{c-i\infty}^{c+i\infty} \frac{d\theta}{2\pi i}
 \exp(\theta {\cal J}) {\cal Z}(\theta;K),
 \label{Laplace2}
 \end{equation}
where it is assumed that the integration contour, considered in the complex plain of $\theta$, goes on the right from all the singularities (poles and cuts) of ${\cal Z}(\theta;K)$. In the path integral
representation GF gets the following form
 \begin{eqnarray}
 && {\cal Z}(\theta;K)\sim \int {\cal D}x\ {\cal D}p\
 \exp\Biggl(\int_0^T dt \Biggl( -\frac{\theta}{2}x^*\tilde{Q} x
 +p^* (\partial_t x+\mu x) +\frac{1}{2}p^*Vp \Biggr)\Biggr),
 \label{Z-path_int}\\
&& \tilde{Q}=Q+K^*RK,\quad  \mu=BK-A,\label{tildeQ+mu}
 \end{eqnarray}
where $p$ is an auxiliary vector variable (momentum). Here and everywhere below we assume that, even if the dynamics was not stable before application of the control,  control stabilizes it. Formally,  this means that $\mu$,  defined by Eq.~(\ref{tildeQ+mu}), has no eigenvalues with negative real values.
The "boundary" ($F$-dependent term) in Eq.~(\ref{Z-path_int}) was ignored, assuming that (like in the one-dimensional LQ case discussed above) it may only influence how the optimum is approached (backwards in time) but remains inessential for describing asymptotic behavior of the optimal control. This path integral is (most conveniently) evaluated by changing to the Fourier (frequency) domain, expressing pair correlation function as the frequency integral, and then relating it to the derivative of the log-GF over $\theta$,
\begin{eqnarray}
&& \langle x_i x_j\rangle=\int\limits_{-\infty}^{+\infty}\frac{d\omega}{2\pi}
\left(\omega^2 V^{-1}+\mu^*V^{-1}\mu+\theta \tilde{Q}\right)^{-1}_{ij},\label{pair-corr}\\
&& \frac{\partial \log {\cal Z}(\theta;K)}{\partial \theta}=-\frac{T}{2}\langle x^*\tilde{Q}x\rangle.
\label{log-derivative}
\end{eqnarray}
Here in Eqs.~(\ref{pair-corr},\ref{log-derivative}) the averaging is over the path integral measure described by Eq.~(\ref{Z-path_int}).  Further, evaluating the integral over $\theta$,  fixing normalization, ${\cal Z}(0;K)=1$, and using the standard formula of matrix calculus, $d/d\theta \log\det(\theta\cdot1+D)=\mbox{tr}((\theta\cdot 1+D)^{-1})$, where $1$ stands for the unit matrix, one arrives at
the following expression
\begin{eqnarray}
\label{Z-full}\log ({\cal Z}(\theta;K))=-\frac{T}{2}\int\limits_{-\infty}^{+\infty}\frac{d\omega}{2\pi}
\log\frac{\det\left(\omega^2 V^{-1}+\mu^*V^{-1}\mu +\theta \tilde{Q}\right)}{\det\left(\omega^2 V^{-1}+\mu^*V^{-1}\mu\right)},
\end{eqnarray}
which is asymptotically exact at $T\to \infty$. Moreover, one can show that for any (spatially) finite system
next order corrections to the rhs of Eq.~(\ref{Z-full}) are $O(1)$.
Note that this representation (\ref{Z-full}) of the log-GF, as an integral over frequency of a log-det, is similar to the relation discussed in Section 3 of \cite{Cb1988} in the context of linking the RS-LQG control to the maximum entropy formulation of the $H_\infty$ control. The log-det has also appeared in \cite{07TCCP} where statistics of currents were analyzed in general non-equilibrium (off-detailed-balance) linear system.

To gain intuition let us first analyze Eq.~(\ref{Z-full}) in  the simple scalar case where the integral on the rhs can be evaluated analytically
\begin{eqnarray}
\log ({\cal Z}(\theta;K)) =\frac{T}{2}
 \left(\mu-\sqrt{\mu^2+\theta \tilde{Q} V}\right).
\label{Z-full-1d}
\end{eqnarray}
Substituting this expression into Eq.~(\ref{Laplace2}) and estimating the integral over $\theta$ in a saddle-point approximation (justified when $T$ is large) one arrives at the LD expression (\ref{LD}) where
\begin{eqnarray}
{\cal S}_*(j)=\frac{V(Q+RK^2)}{16 j}+\frac{(BK-A)^2j}{V(Q+RK^2)}-\frac{BK-A}{2}.
 \label{pdfJ}
 \end{eqnarray}
The LD function is obviously convex and it is defined only for positive $j$. (The asterisk marks,  as before,  that the average and the probability are computed conditioned to yet unspecified $K$.) ${\cal S}_*(j)$ achieves its minimum at, $\langle j \rangle_* =-T^{-1}\left. \partial \log{\cal Z}/\partial\theta\right|_{\theta=0}
=V\tilde{Q}/(4\mu)$, and shows linear asymptotic, ${\cal S}_*(j)\approx j \mu^2/(V\tilde{Q})$, at $j\gg\langle j\rangle$.  Note,  that the aforementioned asymptotic is associated with the cut-singularity in the complex $\theta$ plane of the GF expression (\ref{Z-full-1d}). Indeed,  substituting Eq.~(\ref{Z-full-1d}) into Eq.~(\ref{Laplace2}) and shifting the integration contour to the left, thus forcing it to surround anti-clockwise the $]-\infty;\theta_*=-\mu^2/(V\tilde{Q})]$ cut,  and then estimating the integral by a small part of the contour surrounding vicinity of the cut tip at $\theta_*$, we arrive at the aforementioned $j\gg\langle j\rangle$ asymptotic, ${\cal S}(j)\approx -j\theta_*$.

Returning back to analysis of the general formulas (\ref{Z-full},\ref{Laplace2}), one observes that even though to reconstructing $S_*(j)$ in its full integrity explicitly as a function of $K$ does not look feasible,  we can still, motivated by the scalar case analysis, make some useful general statements about both the average, $\langle j\rangle_*$, and the $j\gg\langle j\rangle_*$ asymptotic of ${\cal S}_*(j)$. We will start from the latter problem.

For analysis of the tail the key object of interest is the $\det$ in Eq.~(\ref{Z-full}) considered at zero frequency, $\omega=0$. Specifically,  one aims to find the zero of the determinant with the largest real value:
\begin{eqnarray}
\theta_*=\max\limits_\theta \mbox{Re}\left(\theta\right)_{
\det\left(\mu^*V^{-1}\mu +\theta \tilde{Q}\right)=0}.
\label{max_zero}
\end{eqnarray}
Indeed,  any zero (there might be many of these in the general matrix case) marks the tip of the respective cut singularity of ${\cal Z}(\theta;K)$ in the complex $\theta$-plane.  Then, the tail, $j\gg\langle j\rangle_*$, asymptotic of the LD function becomes, ${\cal S}_*(j)=-j\theta_*$. Note,  that this linear in $j$ estimation is valid only in the case of a finite system,  when the set (spectrum) of zeros (defined by the condition in Eq.~(\ref{max_zero}) is discrete. In the case of an infinite system,  when the spectrum of zeros becomes quasi-continuous, one needs to account for the multiple zeros, as illustrated in the "string" example of Section \ref{sec:String}.

To evaluate $\langle j\rangle_*$ (as a function of $K$) in the general case one first analyzes it in the time representation. Substituting the $u=-Kx$ ansatz with constant $K$ in Eq.~(\ref{stoch}), expressing $x(t)$ formally as an integral over time (for a given realization of the noise), substituting the result into Eq.~(\ref{cost}), averaging over noise, and then taking the $T\to\infty$ limit one arrives at
\begin{eqnarray}
\langle j\rangle_*=
\frac{1}{2}\int\limits_0^\infty\mbox{tr}\left(Ve^{-\mu^*t}\tilde{Q}e^{-\mu t}\right) dt
=\frac{1}{2}\left.\mbox{tr}\left(V\Pi\right)\right|_{\mu^*\Pi+\Pi\mu=\tilde{Q}},
\label{j-average-time}
\end{eqnarray}
where the latter expression is implicit (as the condition is a matrix one, thus not resolvable explicitly in general) function of $K$. It is straightforward but tedious to check (introducing matrix Lagrangian multiplier for the condition in Eq.~(\ref{j-average-time}) and making variation over $K$ and $\Pi$) that optimization of Eq.~(\ref{j-average-time}) over $K$ results in the algebraic Riccatti equation equivalent to Eq.~(\ref{Pi}) with the first term ignored. Note that the fact that the optimal control derived from the optimization of the average cost function in the stochastic case coincides with the result of the deterministic optimization (ignoring stochasticity) is the fact very well known in the control theory. \footnote{ This fact is akin to a similar statistical physics statement naturally emerging in analysis of the linear stochastic systems driven by white-Gaussian noise and characterized by flux/current (see e.g. \cite{07TCCP}): the average flux corresponds to the minimum of a functional quadratic in the flux.}
The optimal value of the functional in the deterministic case saturates to a constant at $T\to\infty$, while in the stochastic case the average optimal cost grows with time linearly. Asymptotic convergence of the two seemingly different schemes to the same optimal control is thus an indication of the asymptotic self-consistency of the linear ansatz (\ref{u-opt}).

Differentiating Eq.~(\ref{Z-full}) over $\theta$ and then setting $\theta$ to zero,  one derives an alternative (to Eq.~(\ref{j-average-time})) representation for the average rate of the cost function conditioned to $K$
\begin{eqnarray}
\langle j\rangle_*=\int\limits_{-\infty}^{+\infty}\frac{d\omega}{4\pi}\mbox{tr}\left(\left(\omega^2V^{-1}+
\mu^*V^{-1}\mu\right)^{-1}\tilde{Q}\right).
\label{j-average-frequency}
\end{eqnarray}
Note that comparison of Eqs.~(\ref{Z-full},\ref{j-average-time},\ref{j-average-frequency}) also allows to derive expression for the derivative of the log-GF as a time integral, and then have it presented in an implicit algebraic form
\begin{eqnarray}
\label{der-log-Z1}
-T^{-1}\partial_\theta\log {\cal Z}(\theta;K)
&=&\int\limits_{-\infty}^{+\infty}\frac{d\omega}{4\pi}
\mbox{tr}\left(\left(\omega^2V^{-1}+
\mu^*V^{-1}\mu+\theta\tilde{Q}\right)^{-1}\tilde{Q}\right)\nonumber
\\
&=&\frac{1}{2}\mbox{tr}\left(\tilde{V}\Pi\right)_{
\mu^*\Pi+\Pi\mu=\tilde{Q}}
\label{der-log-Z2}\\
&=&\frac{1}{2}\int\limits_0^\infty\mbox{tr}\left(\tilde{V}e^{-\mu^*t}\tilde{Q}e^{-\mu t}\right) dt,
\label{der-log-Z3}
\end{eqnarray}
where $\tilde{V}=V(1+\theta V(\mu^*)^{-1}\tilde{Q}\mu^{-1})^{-1}$.

\section{Optimal Asymptotic Controls}
\label{sec:Opt}

In this Section we formulate the RS-LQ, TO-LQ and CC-LQ asymptotic schemes in the general vector/matrix form as an optimization over $K$. (Note that the asymptotic LQ scheme was already stated as a minimum of Eq.~( \ref{j-average-time}), or equivalently of Eq.~(\ref{j-average-frequency}) in the preceding Section.) Then we illustrate these formulations on the scalar example.

$K_{\mbox{RS}}$, which is asymptotically optimal for the RS-LQ control considered at $\theta>0$, is found by maximizing ${\cal Z}(\theta;K)$. Using Eq.~(\ref{Z-full}) one derives
\begin{eqnarray}
\min\limits_K \left.\int\limits_{-\infty}^{+\infty}d\omega
\log\frac{\det\left(\omega^2 V^{-1}+(BK-A)^*V^{-1}(BK-A) +\theta (Q+K^*RK)\right)}{\det\left(\omega^2 V^{-1}+(BK-A)^*V^{-1}(BK-A)\right)}\right|_{\mbox{Re}(\lambda(BK-A))>0},\label{RS-opt}
\end{eqnarray}
where $\mbox{Re}(\lambda(BK-A))>0$ denotes the stability condition ensuring that the real values of all the eigen-values of $BK-A$ are positive. Note that constancy of the stationary RS-LQ optimal control was proven in \cite{Jacobson1973}, therefore making our approach self-consistent. An alternative,  but obviously equivalent,  formulation of the RS-LQ optimal control consists in minimizing $-T^{-1}\partial_\theta\log {\cal Z}(\theta;K)$. Going along this path and utilizing Eq.~(\ref{der-log-Z2}) one arrives at
\begin{eqnarray}
\min_{K,\Pi}\frac{1}{2}\mbox{tr}\left(V(1+\theta V(\mu^*)^{-1}\tilde{Q}\mu^{-1})^{-1}\Pi\right)_{
\mu^*\Pi+\Pi\mu=Q+K^*RK},
\label{RS-LQ-simple}
\end{eqnarray}
generalizing the LQ formulation stated in the preceding Section as the minimization of Eq.~(\ref{j-average-time}). Solving Eq.~(\ref{RS-LQ-simple}) is reduced to analysis of the respective generalization of the Riccati equations  which can than be turned into a linear eigen-value problem described within the so-called Hamiltonian approach to the RS-LQ problem discussed in \cite{Whittle1986}.

From Eq.~(\ref{TO-LQ}), and assuming time-independence of the control, one can state the general asymptotic TO-LQ optimum utilizing Eqs.~(\ref{Laplace2},\ref{Z-full}) as an optimization of a double integral over frequency and $\theta$. However,  in practice one is interested to discuss the TO-LQ optimization only at sufficiently large values of the cost, $j T$. Using analysis of the preceding Section one derives the desired double asymptotic (valid at large $T$ and large $j$) and simpler to state expression describing $K_{\mbox{TO}}$
\begin{eqnarray}
\min\limits_K \max\limits_\theta \mbox{Re}\left(\theta\right)_{\begin{array}{c}\mbox{Re}(\lambda(BK-A))>0\\
\det\left(\mu^*V^{-1}\mu +\theta \tilde{Q}\right)=0\end{array}},
\label{TO-opt}
\end{eqnarray}
where $\max$ is over complex $\theta$ and the optimal LD value of the PDF tail is exponential,
\begin{eqnarray}
\log{\cal P}_{\mbox{TO}}({\cal J})\approx -\mbox{Re}(\theta_{\mbox{TO}}){\cal J},
\label{LD-TO}
\end{eqnarray}
with  $\theta_{\mbox{TO}}$ solving Eq.~(\ref{max_zero}). Note that the $\det=0$ condition in Eq.~(\ref{TO-opt}) is reminiscent of the $\mu$-measure which is the key element of the robust control approach, see \cite{96ZDG,98Zho} and references therein.

In the same double asymptotic (large $T$ and large $j$) regime the optimal CC-LQ control (\ref{CC-LQ}) is given by
\begin{eqnarray}
\left.\min\limits_K \mbox{tr}\left(V\Pi\right)\right|_{\begin{array}{c}\mbox{Re}(\lambda(BK-A))>0\\
\mu^*\Pi+\Pi\mu=\tilde{Q},\\
\max\limits_\theta (\mbox{Re}(\theta))_{\det(\mu^*V^{-1}\mu+\theta\tilde{Q})=0}\geq\frac{\log(1/\varepsilon)}{jT}.\end{array}}
\label{CC-opt}
\end{eqnarray}
Note that unlike Eqs.~(\ref{RS-opt},\ref{TO-opt}), Eq.~(\ref{CC-opt}) does not have valid solutions for any value of the $\log(1/\varepsilon)/{\cal J}$ ratio.  In fact,  it is clear from Eq.~(\ref{LD-TO}) that to have a nonempty solution of Eq.~(\ref{CC-opt}) one needs to require that $\mbox{Re}(\theta_{\mbox{TO}})jT\leq \log(1/\varepsilon)$.
Once the optimum solution is found, one estimates the LD asymptotic of the cost function PDF by an expression similar to the one given by Eq.~(\ref{LD-TO}), with $\mbox{TO}$ subscript replaced by the $\mbox{CC}$ one.

\subsection{Scalar case}
In the remainder of this Section we illustrate all of the aforementioned formulas on the scalar example. In this simple case
integral  on the rhs of Eq.~( \ref{RS-opt})  is equal to
\begin{equation}
2\pi\left(\sqrt{(BK-A)^2+V\theta(Q+RK^2)}-\left(BK-A\right)\right),
\label{RS-opt-int-scalar}
\end{equation}
resulting in the following optimal value
 \begin{equation}
 K_\theta=\frac{A/B+\sqrt{{A^2}/{B^2}+{Q}/{R}+QV\theta/B^2}}{1+VR\theta/B^2}.
 \label{K-theta-scalar}
 \end{equation}

The large deviation tail of the PDF of $j$ at a given $K$ can be extracted from Eq.~(\ref{pdfJ}):
  \begin{equation}
 j\gg \langle j\rangle:\quad {\cal S}_*(j)\approx \frac{(BK-A)^2}{V(Q+RK^2)} j +o(j).
 \label{coeff-j}
 \end{equation}
Optimizing the PDF over $K$ we find two different cases depending  on the sign of $A$. At $A>0$ coefficient in front of the linear in $j$ term on the rhs of Eq.~(\ref{coeff-j}) grows monotonically with $K$ from the $(A/B,+\infty)$ interval.
To find the optimal value of $K$ in this case one has to take into the $O(j)$ term thus deriving :
 \begin{eqnarray}
A>0:\quad K_{\mbox{TO}}=\sqrt{4Aj}{RV},\quad \log{\cal P}_{\mbox{TO}}\approx -\frac{B^2jT}{RV}.
\label{TO-scalar}
\end{eqnarray}
In the other case of $A=-|A|<0$ the linear coefficient in Eq.~(\ref{coeff-j}) reaches its maximum at $K=BQ/(R|A|)$, thus
resulting in
 \begin{eqnarray}
A<0:\quad \log{\cal P}_{\mbox{TO}}\approx -\frac{jT}{RVQ}\left(RA^2+B^2Q\right).
\label{TO-scalar1}
\end{eqnarray}

Finally,  the CC-optimal formula (\ref{CC-opt}) has no solution if $B^2j/(RV)<c$ in the $A>0$ case and
if $j\left(RA^2+B^2Q\right)/(RVQ)<c$ in the  $A<0$ case. (Here we assume, as above, that $\epsilon(0;T)=\exp(-c T)$.)
When $\epsilon$ is chosen sufficiently small (i.e. $c$ is sufficiently large), the feasibility domain in Eq.~(\ref{CC-opt}) is not
empty and one distinguishes two regimes depending on how $K_\varepsilon$, defined by
\begin{eqnarray}
K_\varepsilon=\frac{1}{1-\kappa}\left(\frac{A}{B}+\sqrt{\kappa\left(\frac{A^2}{B^2}+(1-\kappa)\frac{Q}{R}\right)}\right),
\quad \kappa=\frac{cVR}{B^2j},
\label{K-epsilon-scalar}
\end{eqnarray}
compares with $K_0$, which is the bare LQ optimal value correspondent to $K_\theta$ from Eq.~(\ref{K-theta-scalar}) evaluated at $\theta=0$. One derives
\begin{equation}
K_{\mbox{CC}}=\max\left(K_\varepsilon, K_0\right),
\label{CC-scalar}
\end{equation}
where of the two regimes one is achieved within the interior of the optimization domain (tail constraint is not restrictive)
while the other one corresponds to the tail imposed by the boundary of the domain. It is worth noting that
(\ref{K-epsilon-scalar}) is valid for both signs of $A$.

\section{Example of a String}
\label{sec:String}

In this Section we discuss an explicitly solvable example of an infinite system where the set of zeros (of the determinant in the condition of Eq.~(\ref{max_zero})) forms a quasi-continuous spectrum.
Consider a string, defined as an over-damped system of multiple bids on a line connected to each other by elastic springs of strength $D$, stretched by the linear force of the strength $A$ and subject to Langevien driving:
 \begin{eqnarray}
 && \partial_t x_j=Ax_j +D(x_{j+1}+x_{j-1}-2x_j)
 +B u_j +\xi_j,
 \label{spectr1} \\
 && J=\frac{1}{2}\int_0^T dt\ \sum_j (Qx_j^2+R u_j^2),
 \label{spectr2}
 \end{eqnarray}
where $j=1,\cdots,N$, $x_j$ marks position of the $j$-th bid of the string, and the zero-mean white-Gaussian noise is
distributed as in Eq.~(\ref{xi}) with $V_{ij} = V\delta_{ij}$. $u_j$ in Eq.~(\ref{spectr1}) stands for control. We are looking for a time-independent linear in $x$ control,  assuming that the control acts uniformly on all bids of the string, i.e.   $u_j= - K x_j$. Let us also assume that the string is periodic with the period $N$. Then, solution of Eq.~(\ref{spectr1}) allows expansion in the series over spatial harmonics
 \begin{equation}
 x_j=\sum_{j=1}^N \exp(i q(j/N)) x_q,
 \label{spectr5}
 \end{equation}
with the wave vector, $q$, from the interval, $-\pi<q<\pi$, and  resulting in the following separated equations for the individual harmonics
 \begin{equation}
 \partial_t x_q=Ax_q -2D(1-\cos q) x_q
 - B K x_q +\xi_q.
 \label{spectr6}
 \end{equation}

Repeating the steps leading to (\ref{Z-full-1d}) one arrives at
 \begin{eqnarray}
 \log {\cal Z}=\frac{T}{2}\sum_q\left(
 BK-A +2D(1-\cos q)-\sqrt{(BK-A+2D(1-\cos q))^2+V(Q+RK^2)\theta}\right).
 \label{spectr9}
 \end{eqnarray}
We choose to analyze only the most interesting regime, $D\gg BK- A$, when a nontrivial collective behavior emerges. Then, in the long wave-length, $1-\cos q \to q^2/2$, and continuous, $\sum_q\to (N/2\pi) \int dq$, limits one derives
 \begin{eqnarray}
 && \log {\cal Z}=-\frac{TN}{3\pi}
 \frac{(BK-A)^{3/2}}{D^{1/2}}
 (1+s)^{1/4} i
 \left((1+\sqrt{1+s})
 {\cal K}\left(\frac{1}{2}-\frac{1}{2\sqrt{1+s}}\right)
 -2 {\cal E}\left(\frac{1}{2}
 -\frac{1}{2\sqrt{1+s}}\right) \right),
 \label{spectr10} \\ &&
 s=\frac{V(Q+RK^2)\theta}{(BK-A)^2},
 \label{spectr11}
 \end{eqnarray}
where one utilizes the standard ${\cal K},{\cal E}$ notations for the elliptic functions.

Expression on the rhs of Eq.~(\ref{spectr10}) shows  a singularity at $s=-1$, coinciding with the singularity (in the complex $\theta$ plane) observed in the scalar case at $\theta_*$. Substituting Eq.~(\ref{spectr10}) into Eq.~(\ref{Laplace2}) and evaluating the integral over $\theta$  in the saddle-point approximation one arrives at
 \begin{eqnarray}
{\cal S}_*(j)
 \approx\frac{\sqrt2\ N(BK-A)^{3/2}}{3\pi D^{1/2}}
 +\theta_* j.
 \label{spectr13}
 \end{eqnarray}
Juxtaposing the string expression Eq.~(\ref{spectr13}) to the scalar one Eq.~(\ref{pdfJ}) one notes different behaviors with respect to $BK-A$. Optimizing Eq.~(\ref{spectr13}) over $K$ at a given large value $J=jT$,
one obtains the same tail expression, second formula in  (\ref{TO-scalar}),  however with another optimal control
 \begin{equation}
 K^{5/2}_{\mbox{str}}=\frac{2^{3/2}\pi A D^{1/2}}{RVTN B^{1/2}} J,
 \label{spectr14}
 \end{equation}
replacing the first formula in Eq.~(\ref{TO-scalar}). Note that in the string case the optimal $K$ scales as $J^{2/5}$ which should be contrasted to the $J^{1/2}$ scaling in the scalar case from Eq.~(\ref{TO-scalar}).

\section{Conclusions and Path Forward}
\label{sec:Con}

This manuscript contributes the subject in control theory - designing control scheme with some guarantees not only on the average of the cost functions but also on fluctuations,  specifically extreme fluctuations related to the tail of the cost function PDF. We consider linear, first order in time derivative, stochastic system of the Langevien type subject to minimization of a quadratic cost function and also with (chance) constraints imposed on the tail of the cost function PDF. In the stationary regime of large time,  when control is sufficient to make the system stable, we reduce the stochastic dynamic problem of the "field theory" type to static optimization analysis with objectives and constraints stated in a matrix form. This type of reduction is unusual in the system lacking the fine-tuned Fluctuation Dissipation relation between relaxational and stochastic terms. On the other hand, the progress made is linked to linearity of the underlying stochastic systems which allowed, as in some problems of passive scalar turbulence \cite{95CFKL,98CFK,01FGV} and driven linear-elastic systems \cite{05KAT,07TCCP}, to formally express solution for the system trajectory as an explicit function of the noise realization. Besides that, main technical ingredients, which allowed us to derive the results, consisted in making plausible assumption about the structure of the control (linear in the state variable and frozen in time),  and then performing asymptotic evaluations of the cost functions statistics conditioned to the value of the cost matrix. Techniques of path integral,  spectral analysis and large deviation estimations were used. We tested results on the simple scalar case and illustrated utility of the method on an exemplary high-dimensional system (1d chain of particles connected in a string).

We plan to continue exploring the interface between control theory and statistical physics addressing the following challenges.
\begin{itemize}
\item Computational feasibility of the main formulas of the paper, stating  RS-, TO- and CC- controls in Eqs.~(\ref{RS-opt},\ref{TO-opt},\ref{CC-opt}) as static optimization problems, need to be analyzed for large systems and networks.   After all main efforts in the applied control theory go into designing efficient algorithms for discovering optimal,  or close to optimal,  control,  and we do plan to contribute this important task. Therefore, further analysis is required to answer the important practical question: if the static formulations of the newly introduced TO-QG and CC-QG controls allow computationally favorable exact or approximate expressions in terms of convex optimizations?

\item We also plan to study weakly non-linear stochastic systems through a singular perturbation stochastic diagrammatic technique of the Martin-Siggia-Rose type \cite{73MSR}. Besides, some of the methods we used in the manuscript, especially related to large deviation analysis, are not restricted to linear systems. Our preliminary tests show that effects of the non-linearity on the PDF tail are seriously enhanced in comparison with how the same nonlinearity influences the average case control.

\item It will be interesting to study TO- and CC- versions of the path-integral nonlinear control problems discussed in \cite{05Kap,11Kap,10BWK,Dj11}. These problems,  in their standard min-cost formulations,  allow reduction (under some Fluctuation-Dissipation-Theorem like relations between the form of control, covariance matrix of the noise and the cost function) from the generally non-linear Hamilton-Jacobi-Bellman equations for the optimal cost function to a linear equation of a Schr\"{o}dinger type.

\item The effects of partial observability and noise in the observations can be easily incorporated in both TO- and CC- schemes discussed in the paper. In fact this type of generalization is standard and widespread in the control theory,  where for example the LQG (Linear-Quadratic-Gaussian) control generalizes the LQ control.

\item In terms of relevance to an application, this work was motivated by recent interest and discussions related to developing new optimization and control paradigms for power networks,  so-called smart grids. In this application, strong fluctuations associated with loads and renewable generation, electro-mechanical control of generation, desire to make the energy production cheaper while also (and most importantly) maintaining  probabilistic security limitations of the chance-constrained type - all of the above  make the theoretical model discussed in this paper an ideal framework to consider.  In particular, we plan to extend the approaches of \cite{11DBC,12HBC} and modify and apply the theory developed in this manuscript to design a multi-objective Chance Constrained Optimum Power Flow including better control of generation, loads and storage resources in power grids.

\item We also anticipate that some of the models and results discussed in the paper are of interest for problems in statistical micro- and bio- fluidics, focusing on adjusting characteristics of individual molecules (polymers, membranes, etc) and also aimed at modifying properties of the medium (non-Newtonian flows) macroscopically. Time independent and linear nature of the control schemes discussed in the paper make them especially attractive for these applications. Natural constrains, e.g. associated with the force-field (optical or mechanical) as well as with some other physical limitations, could be incorporated into control as single- or multi-objective cost functions.

\end{itemize}

We are thankful to D. Bienstock, L. Gurvits, H.J. Kappen, K. Turitsyn and participants of the "Optimization and Control Theory for Smart Grids" project at LANL for motivating discussions and remarks. Research at LANL was carried out under the auspices of the National Nuclear Security Administration of the U.S. Department of Energy at Los Alamos National Laboratory under Contract No. DE C52-06NA25396.

\bibliographystyle{unsrt}
\bibliography{control,path_int}

\end{document}